\documentclass[12pt]{article}
\title{Kolmogorov and von Mises viewpoints to the 
Greenburger-Horne-Zeilinger paradox}
\author{Andrei Khrennikov\\
Department of Mathematics, Statistics and Computer Sciences\\
University of V\"axj\"o, S-35195, Sweden}
\begin{document}
\maketitle

\begin{abstract} We present comparative probabilistic analysis of the Greenburger-Horne-Zeilinger paradox
in the frameworks of Kolmogorov's (measure-theoretical) and von Mises' (frequency) models of the
probability theory.  This analysis demonstrated that the GHZ paradox is merely a consequence 
of the use of Kolmogorov's probabilistic model. By using von Mises' frequency approach we 
escape the contradiction between the local realism and quantum formalism. The frequency approach
implies automatically contextual interpretation of quantum formalism: different collectives induce different probability distributions.
On the other hand, the formal use of Kolmogorov's model implies the identification of 
such distributions with one abstract Kolmogorov measure. In the measure-theoretical approach
we can escape the paradox, if we do not suppose that probability distributions
corresponding to different settings of measurement devices are equivalent. We discuss the connection
between equivalence/singularity dichotomy in measure theory and the existence of compatible and noncompatible
observables.
\end{abstract}

\section{Introduction}

It is well known that violations of Bell's inequality [1] by
quantum correlations in the Einstein-Podolsky-Rosen (EPR)
framework may be interpreted as the evidence of the impossibility
to use the {\it local realism} in quantum theory (see, for example, [2], [3]).
Such a viewpoint was strongly supported by experiments of Aspect
[2] which demonstrated violations of generalized Bell's inequality
(see also [3]). Despite of the general attitude to connect violations of
Bell's inequality with such problems as {\it determinism and locality},
there exists sufficiently strong opposition [4]-[9] to 
such a conclusion. This opposition, despite of
the great diversity of approaches, can be called the {\it probability
opposition.} The general viewpoint of adherents of the
probabilistic interpretation of violations of Bell's inequality is
that the derivation of this inequality is based on numerous
(hidden) probabilistic assumptions. Unfortunately at the present
time there are no experimental facts which can justify 
these probabilistic assumptions. 
It seems that theoretical as well as experimental
investigations of the EPR paradox (in particular, Bell's
inequality) must be at least partly reoriented to the
investigation of probabilistic roots of this paradox. 

The viewpoint that the notion of probability plays the large role 
in Bell's (and in  EPR's) considerations  is not so new, [4]-[9]. The main consequence of
all these probabilistic analyses is that the EPR experiment could not be described
(as it was assumed by J. Bell, [1]) by the unique
Kolmogorov probability distribution. In fact, these are 
various forms of contextual interpretation of quantum formalism): 
(1) De Broglie, Lochak, Nelson, 
De Muynck, De Baere, Marten, Stekelenborg,  [4], thermodynamical approach to Bell's problem,
difference between hidden and observed probabilities; 
(2) Beltrametti and Cassinelli [5], quantum logic;
(3) Accardi [6], quantum probabilities, no Bayes'
formula; (4) Pitowsky and Gudder [7], probability manifolds; (5) De Baere [7], fluctuating probabilities;
(6) Fine and Rastal [7], no simultaneous probability distribution; (7) Muckenheim [7], negative probabilities;
(8) Khrennikov [8], $p$-adic probabilities; fluctuating probabilities and modified Bell's inequality [9].

However, a new strong argument in the favour of
nonlocal (or nonreal) interpretation of the EPR paradox [10] was
given by so called Greenberger-Horne-Zeilinger (GHZ) paradox, [11].
The GHZ scheme is based on the probability one arguments. From the
first point of view all probabilistic circumstances of the GHZ
scheme are so straightforward that there is no more place for probabilistic
counter arguments. However, the careful probabilistic analysis 
demonstrates that the GHZ paradox has even deeper connection to
foundations of probability theory than Bell's inequality. Roughly
speaking the root of the GHZ paradox might be in the use of the
conventional probability calculus, namely Kolmogorov's (axiomatic)
measure theoretical approach, 1933, [12].

In this paper we shall consider the GHZ paradox from the viewpoint
of so called frequency probability theory, R. von Mises, 1919 [13]
(see [14] for the advanced formalism). In the opposite to Kolmogorov's
model of probability theory which is characterized by the highest
degree of abstraction, von Mises' model of probability theory is
characterized by its concreteness. By R. von Mises we cannot
consider a probability distributions without the relation to the
concrete collective (random sequence). Von Mises' slogan was:
"first
collective and then probability distribution". Analysis of the GHZ
paradox based on von Mises' approach demonstrated that it is rather
doubtful that there exists a collective which produces the
probability distribution which is formally (via Kolmogorov's
approach) used in the GHZ considerations. Hence if we use a
mathematical model of probability theory which is different from
Kolmogorov's model, namely von Mises' model, we observe no
paradox in the GHZ considerations. In particular, there 
is no contradiction between the quantum formalism with the
frequency interpretation of probability and local realism.

Of course, our probabilistic considerations could not be
considered as arguments in favour of either locality or determinism.
It may be that physical reality is nonlocal or even nonreal.
However, Bell's as well as GHZ's approaches do not give definite
arguments to deny locality or determinism. Both these
approaches
are strongly based on the use of one particular model of
probability theory, Kolmogorov's model.

Our frequency analysis clarifies the measure-theoretical roots of the
GHZ paradox. In fact, this paradox can be escaped even in the measure-theoretical
approach if it would not be assumed that probability distributions
corresponding to different settings of measurement devices are equivalent
measures. We discuss the connection
between equivalence/singularity dichotomy in measure theory and the existence of compatible and noncompatible
observables. It seems that the splitting of physical reality to classical and quantum realities is just
a consequence of the general (mathematical) property of probability measures. So 
it is just a property of our (mathematical) description of physical reality.

The Kolmogorov definition of a probability space is well known [12], [15].
This is a triple ($\Omega, F, {\bf P}$), where $\Omega$ is an
abstract
set, $F$ is a $\sigma$-field of subsets (events) of $\Omega,
{\bf P}$ is the probability (normalized by 1 and $\sigma$-additive)
measure on $F.$ On the other hand, the frequency probability
theory
of R. von Mises is now days practically forgotten. So we must
present an extended introduction to this approach, see section 2.

We must remark that von Mises' theory was strongly criticized due
to rather informal definition of randomness, [16]. In fact, this
purely mathematical critique was one of the reasons to eliminate
the frequency approach from quantum formalism. We do not relate
our use of frequency theory to sophisticated mathematical
problems
of randomness [16]. There are two main reasons to eliminate the
problem of randomness from physical considerations and justify the
use of the frequency formalism. The first is von Mises' 
observation that the class of place selections must be determined
not by some mathematical theory, but by the concrete physical
phenomenon. This viewpoint is supported by Wald's theorem [17] by
that if we fix a countable set of place selections, then there
exist sufficiently many collectives with respect to this set of
place selections.
The second is my own observation that it seems to be that the
property of randomness is not related (at least directly) to
physical measurements (at least for present experiments). We are
always interested only in one property of a sequence of
observations: the statistical stabilization of relative frequencies
$\nu_N={\frac{n}{N}}$ to some limiting quantities ${\bf P}$
(probabilities).

\section{Frequency probability theory}

{\bf 2.1. History.} {\small The frequency probability theory was developed by R. von Mises in 1919 (see [13], [14], [9]
for the details).
In fact, the basis of the frequency approach was provided in the work of J. Venn, 1866, see [18].
The frequency theory was used as the motivation of Kolmogorov's axiomatic, 1933, of 
the conventional probability theory (see remarks in [12]). The main advantage of the conventional theory 
is its {\it abstractness.} Here we work with abstract probability distributions which are not directly
related to the concrete physical model. Thus results of the conventional probability theory can be used
without any modification in any physical models. However, this advantage may become in some circumstances
a disadvantage, because the abstractness of the formalism does not give the possibility to analyse
the origin (and even the existence) of probability distributions. On the other hand, the frequency theory
of probability is concrete. Here to introduce a probability distribution, we must be sure that there
exists a collective (random sequence) which produces this probability distribution. The collective is more primary object
than a probability distribution. The collective has more direct connection with a physical phenomenon.
However, in the frequency approach we cannot obtain results which are valid for `all probability distributions'.
The probability distribution without a collective is nothing. Typically such a concreteness 
is considered as the large disadvantage of the frequency approach (comparing with the conventional measure
theoretical approach). Of course, it is more attractive to prove some probabilistic statement ones
and then to apply it to numerous physical models. This was one of the reasons to eliminate the frequency approach
from applications in the favour of the measure-theoretical approach. 
\footnote{Another reason was the problem of the rigorous mathematical definition of 
a collective, random sequence, see, for example, [9].} 

In the present paper we demonstrate that the frequency analysis of probabilistic assumptions
for the derivation of Bell's inequality can give some new sights to this problem.
These sights would be impossible
to obtain in the conventional abstract framework. Analysis of collectives can give more than
analysis of abstract probability distributions.}

{\bf 2.2. Collective.} Let ${\cal E}$ be an ensemble of physical systems. We take elements 
of ${\cal E}$ and form a sequence
$
\pi= (\pi_1,\pi_2,..., \pi_N,... )\;.
$
Suppose that elements of ${\cal E}$ have some properties.
\footnote{It is not important in general either these properties are objective (properties of an object)
or `created' in the process of observation by an observer, see [3].} Suppose that these
properties can be described by  natural numbers, $ L=\{1,2,...,m \}$
(the set of `labels'). In principle we can consider
continuous label sets, see [14]. Thus, for each $\pi_j\in \pi,$ we have a number $\alpha_j\in L.$ 
So $\pi$ induces a sequence 
\begin{equation}
\label{la}
x=(\alpha_1,\alpha_2,..., \alpha_N,...), \; \; \alpha_j \in L.
\end{equation}
For each fixed $\alpha \in L,$ we have the relative frequency
$\nu_N(\alpha)= n_N(\alpha)/N$ of the appearance of $\alpha$
in $(\alpha_1,\alpha_2,..., \alpha_N).$ 

R. von Mises said that
$x$ satisfies to the principle of the {\it statistical stabilization} of relative frequencies,
if, for each fixed $\alpha \in L,$ $\vert \nu_N(\alpha)- \nu_M(\alpha)\vert\to 0, N, M \to \infty.$
The corresponding limit 
\begin{equation}
\label{l0}
{\bf p} (\alpha)=\lim_{N\to \infty} \nu_N(\alpha)
\end{equation}
is said to be a probability.
This probability can be extended to the field of all subsets of $L:$
\begin{equation}
\label{l1}
{\bf p} (B)= \lim_{N\to \infty} \nu_N(\alpha\in B)= \lim_{N\to \infty} \sum_{\alpha\in B} \nu_N(\alpha)=
\sum_{\alpha\in B} {\bf p}(\alpha)\; , B \subset L
\end{equation}
(the situation becomes sufficiently complex for an infinite $L,$ see Tornier [11]). We remark that 
${\bf p}(L)=1.$

R. von Mises said that $x$ satisfies the principle of {\it randomness} if limits (\ref{l0}) 
are invariant with respect to choices of some subsequences in $x.$ These choices of subsequences,
so called place selections, have some properties, see [13], [14] or  [9] (which are unimportant for our investigation).
\footnote{The class of place selections was not defined precisely by R. von Mises. This induced
numerous discussions. However, the problem can be solved (at least partially) by the consideration
of countable classes of place selections, Wald theorem, [17] or [9], p.43.} In principle the reader may forget
about the principle of randomness and consider only the principle of the statistical stabilization.
It seems that only this principle is important (at least at the moment) in physics in that we study 
behaviour of frequencies.

Sequence (\ref{la}) which satisfies to two von Mises' principles is said to be a {\it collective};
${\bf p}$ is said to be a {\it probability distribution} of the collective $x.$ We will often use the symbols
${\bf p}(B;x)$ (and $\nu_N(B;x), n_N(B;x)), B\subset L,$ to indicate the dependence on the concrete collective
$x.$

The frequency probability formalism is not a calculus of probabilities. It is
a {\it calculus of collectives.} Thus instead of operations for probabilities
(as it is in the conventional probability theory), we define operations for collectives.

{\bf 2.3. Operation of combining of collectives.} This operation will play the
crucial role in our analysis of probabilistic foundations of Bell's arguments.
Let $x=(x_j)$ and $y=(y_j)$ be two collectives with label
	sets $L_x$ and $L_y$, respectively. We define a new sequence
$
z=(z_j),\ z_j=\{ x_j,y_j\}
$
(in general $z$ is not a collective).
	Let $a\in L_x$ and $b\in L_y$. Among the first $N$ elements of
	$z$ there are $n_N(a;z)$ elements with the first component equal
	to $a$. As $n_N(a;z)=n_N(a;x)$ is a number of $x_j=a$ among the
	first $N$ elements of $x$, we obtain that
	$\lim_{N\to\infty}\frac{n_N(a;z)}{N}={\bf p}(a;x)$. Among these
	$n_N(a;z)$ elements, there are a number, say $n_N(b/a;z)$ whose
	second component is equal to $b$. The frequency $\nu_N(a,b;z)$
	of elements of the sequence $z$ labeled $(a,b)$ will then be
	$$
	\frac{n_N(b/a;z)}{N}=\; \; \frac{n_N(b/a;z)}{n_N(a;z)}\; \; \frac{n_N(a;z)}{N}\;.
	$$
	We set $\nu_N(b/a;z)=\frac{n_N(b/a;z)}{n_N(a;z)}$. Let us assume
	that, for each $a\in L_x$, the subsequence $y(a)$ of $y$ which
	is obtained by choosing $y_j$ such that $x_j=a$ is an
	collective.
	Then, for each $a\in L_x$, $b\in L_y$, there exists
\begin{equation}
\label{l3}
{\bf p}(b/a;z)=\lim_{N\to\infty}\nu_N(b/a;z)=\lim_{N\to\infty}\nu_N(b;y(a))=
{\bf p}(b;y(a)).
\end{equation}
	We have
$
		\sum_{b\in L_2}{\bf p}(b/a;z)=1. 
$
	The existence of ${\bf p}(b/a;z)$ implies the existence of
	${\bf p}(a,b;z)=\lim_{N\to\infty}\nu_N(a,b;z)$.
	Moreover, we have
\begin{equation}
\label{l2}
{\bf p}(a,b;z)={\bf p}(a;x) \; {\bf p}(b/a;z)
\end{equation}
	and
$
		{\bf p}(b/a;z)={\bf p}(a,b;z)/{\bf p}(a;x), 
$
	if ${\bf p}(a;x)\ne 0$.
	We have
$$
		\sum_{a\in L_a}\sum_{b\in L_2}{\bf p}(a,b;z)=1.
$$
	Thus in this case the sequence $z$ is an collective and  the
	probability distribution ${\bf p}(a,b;z)$ well defined. The
	collective $y$ is said to be {\it combinable} with the collective
	$x$. The relation of combining is a symmetric relation on the set of
	pairs of collectives with strictly positive probability
	distributions $({\bf p} >0).$
	
	{\bf 2.4. Independent collectives.} 
	Let $x$ and $y$ be collectives. Suppose that they are combinable. The $y$ is said to be
		independent from $x$ if all collectives $y(a)$, $a\in L_x$, have the same
		probability distribution which coincides with the probability
	distribution ${\bf p}(b;y)$ of $y$. This implies that 
	$$
	{\bf p}(b/a;z)=\lim_{N\to\infty}\nu_N(b/a;z)=\lim_{N\to\infty}\nu_N(b;y(a))={\bf p}(b;y)\; .
	$$
Here the conditional probability ${\bf p}(b/a;z)$ does not depend on $a.$ Hence
	$$
			{\bf p}(a,b;z)={\bf p}(a;x)\; {\bf p}(b;y),\,a\in L_x,\, b\in L_y.
$$

From the physical viewpoint the notion of independent collectives is more natural 
than the notion of independent events in the conventional probability theory. In latter
the relation ${\bf p}(a,b)= {\bf p}(a) {\bf p}(b)$ can hold just occasionally 
(as the result of a game with numbers, see [14] or [9], p.53).

\section{Kolmorogov's viewpoint to the GHZ scheme}

From the probabilistic viewpoint the GHZ experiment can be
described in the following way ( in the Kolmorogov approach). Let
$(\Omega, {\rm{F}}, {\bf P})$ be a Kolmogorov probability space which
describes hidden variables. For each setting ($\phi_1, \phi_2,
\phi_3)$ of phase shifts we define random variables ${\rm{A}}
(\phi_1,\omega),
{\rm{B}}(\phi_2,\omega), {\rm{C}}(\phi_3,\omega)$
corresponding to physical observables
${\rm{A}}(\phi_1), {\rm{B}}(\phi_2), {\rm{C}}(\phi_3)$ (given by
measurements for
photons 1,2,3 respectively, in the triple (1,2,3)). Quantum
formalism predicts that there exist four settings
($\phi_1^i, \phi_2^i, \phi_3^i), i=1,2,3,4$ such that
\begin{equation}
\label{K0}
{\rm{A}}(\phi_1^i,
\omega) {\rm{B}}(\phi_2^i,\omega) {\rm{C}}
(\phi_3^i,\omega)=1,
\end{equation}
\begin{equation}
\label{K1}
\omega \in \Omega_{\rm{i}}^+ \in {\rm{F}},\; \; {\bf P}(\Omega_{\rm{i}}^+)=1,
\; {\rm{i}}=1,2,3\; ;
\end{equation}
\begin{equation}
\label{K2}
{\rm{A}}(\phi_1^4, \omega) {\rm{B}}(\phi_2^4, \omega)
{\rm{C}}(\phi_3^4, \omega)= - 1,
\end{equation}
\begin{equation}
\label{K3}
\omega\in\Omega_4^- \in {\rm{F}},\; \;  {\bf P}(\Omega_4^-)=1\;.
\end{equation}

By using algebraic properties $({\rm{A, B, C}}= \pm 1)$ we obtain 
that 
\begin{equation}
\label{K4}
\Sigma^+=\Omega_1^+\cap\Omega_2^+\cap\Omega_3^+ \subset \Omega_4^+ =\Omega\setminus \Omega_4^-\;.
\end{equation}

The trivial mathematical considerations in Kolmorogov's framework
imply that by (\ref{K1})
\begin{equation}
\label{K5}
{\bf P}(\Sigma^+)=1\;.
\end{equation}
On the other hand, by
(\ref{K3}) and (\ref{K4}) we have 
\begin{equation}
\label{K6}
{\bf P}(\Sigma^+)=0.
\end{equation}

This is the GHZ paradox. The typical conclusion is that we could
not use the  local deterministic description.

From the Kolmorogov viewpoint it seems that all was right in the
GHZ derivation.

\section{Von Mises' viewpoint to the GHZ paradox}

Here we could not start with an abstract probability distribution
of hidden parameters. First we have to define a collective which
produces this distribution. To introduce a collective, we have to
define the label set $L$ of this collective. It is convenient to
use symbol $\Omega$ instead of $L$ (to use formulas of the
previous section). However, it is just the same symbol and nothing
more. Here $\Omega$ has the following structure: $\Omega=\Lambda \times
\Lambda_{1} \times \Lambda_{2} \times \Lambda_{3},$ where $\Lambda$ is the
set of hidden variables for a quantum system (a triple of
photons), $\Lambda_{\rm{j}}, {\rm{j}}=1,2,3,$ are sets of hidden
variables for measurement devices (for $A, B$ and $C,$
respectively). \footnote{To simplify considerations, we assume that all sets of hidden
variables are finite.}

For each setting $\phi_1, \phi_2, \phi_3$ of phase shifts, we
may consider (in the hidden variables framework) a sequence
${\rm{x}}_{\phi_1 \phi_2 \phi_3}=(\omega_1, \omega_2, \ldots, \omega_{\rm{N}}.
\ldots),
\omega_{\rm{j}}=(\lambda_{\rm{j}}, \lambda_{\rm{j}}^1, \lambda_{\rm{j}}
^2, \lambda_{\rm{j}}^3)\in \Omega,$ where $\omega_{\rm{j}}$ is the
configuration of hidden variables for ${\rm{jth}}$ quantum system
$\pi_{\rm{j}}$ (a triple of photons) + three measurement devices at the  instants of measurements
$j=1,2,...$
 
The first question is the following:
{\it Is $x_{\phi_1 \phi_2 \phi_3}$ a collective?} 
We have no experimental reasons to suppose that micro parameters have the
property of the statistical stabilization (as macro parameters).
It may be that the property of the statistical stabilization on
the macro level is just a consequence of the average over huge
ensembles of hidden parameters. Well, suppose that $x_{\phi_1 \phi_2 \phi_3}$ 
is a collective. Thus the frequency probability distribution
\[{\bf P}_{\phi_1 \phi_2 \phi_3}(\lambda=k, \lambda^1=s_1,\lambda^2=
s_2, \lambda^3=s_3)=\]\[\lim_{N\rightarrow\infty}
{\frac{n_N(\lambda=k, \lambda^1=s_1,\lambda^2=s_2, \lambda^3=s_3)}
{N}}\] is well defined.
\footnote{The consideration of hidden variables for measurement
apparatuses  is quite natural from the physical
viewpoint. In fact, it is the hidden variable representation of
Bohr's ideas.}
So, for four different settings $(\phi_1^{\rm{i}},
\phi_2^{\rm{i}}, \phi_3^{\rm{i}})$
of phase shifts we have four collectives $x^{\rm{i}}=
x_{\phi_1^{\rm{i}},
\phi_2^{\rm{i}}, \phi_3^{\rm{i}}},  {\rm{i}}=1, 2, 3, 4,$ with probability
distributions ${\bf P}_i \equiv {\bf P}_{x^{\rm{i}}}.$
By the GHZ scheme we obtain that 
\begin{equation}
\label{C1}
{\bf P}_i(\Omega_{\rm{i}}^+)=1, {\rm{i}}=1,2,
3,\; \mbox{and}\; \; {\bf P}_4(\Omega_4^+)=0.
\end{equation}
Of course, by (\ref{K4}) we obtain
\begin{equation}
\label{C2}
{\bf P}_4(\Sigma^+)=0.
\end{equation}
However, the first three 
equations in (\ref{C1})
do not imply that 
\begin{equation}
\label{C3}
{\bf P}_4(\Sigma^+)=1.
\end{equation}
Hence there is no paradox. To obtain the paradox, we need to obtain (\ref{C3}). 
Thus there must be some special restrictions on collectives (and consequently
probability distributions) which imply (\ref{C3}).
One of such restrictions is that the probability distribution does
not depend on the setting $(\phi_1, \phi_2, \phi_3)$ of phase
shifts:
\begin{equation}
\label{C4}
{\bf P}={\bf P}_{\phi_1, \phi_2, \phi_3}.
\end{equation}
However, such an assumption has no physical justification (compare with [4]-[9]).
First of all we have to assume so called ensemble reproducibility
for hidden variable $\lambda$ (see [7] and [9]): the preparation
procedure for quantum systems must precisely reproduce the
probability distribution of hidden variables in different runs of
the experiment (in particular, for different settings $\phi_1,
\phi_2, \phi_3$). Despite of the common opinion that such a
reproducibility is a natural property of quantum systems
(preparation procedures), at the present stage of experimental
research it is impossible to test this hypothesis. Moreover, the
hypothesis of reproducibility is a form of the postulate on the
completeness of quantum mechanics. By the
hypothesis on reproducibility we have that quantum state $\psi$
uniquely determines all statistical properties of the (ideal infinite)
ensemble of quantum particles described by $\psi.$

So by accepting this hypothesis we turn back (at least indirectly)
to the original discussion of Einstein, Podolsky, Rosen and Bohr
on the completeness of quantum mechanics. In some sense this is
the logical loop, because one of the main aims of J.Bell and his
followers was to transform the EPR polemic on the completeness of
quantum mechanics into polemic on locality and determinism.

{\bf {Remark}} (On the interpretation of a wave function).{\small Of course, all our
previous considerations on the hypothesis of
reproducibility and the completeness of quantum mechanics strongly
depend on the interpretation of a wave function. In fact, we used
so called statistical interpretation of quantum mechanics (see,
for example, L. Ballentine [19]): a wave function gives the description of
statistical properties of an ensemble of quantum particles. Here
the statistical reproducibility of macro properties need not be
based on the statistical reproducibility of micro properties. For
an adherent of the orthodox Copenhagen interpretation (by that the wave
function provides the complete description of an individual
quantum system), there are no doubts in the validity of the
hypothesis of reproducibility.}

However, even if we suppose that there are no ensemble
fluctuations, there are still some doubts in the validity of (\ref{C4}). It is
more natural to think that different settings of apparatuses
produce different distributions of micro states of these
apparatuses (compare with [4]-[9]).

\section{Singularity/equivalence dichotomy and the principle of
complementarity}

Of course, (\ref{C4}) is only a sufficient condition for obtaining the GHZ
paradox. In fact, we need only that 
\begin{equation}
\label{C5}
{\bf P}_{\phi_1 \phi_2 \phi_3} ({\rm{E}})=0
\leftrightarrow {\bf P}_{\phi_1^\prime \phi_2^\prime \phi_3^\prime}
({\rm{E}})=0
\end{equation}
for any two settings
$\phi_1 \phi_2 \phi_3$ and $\phi_1^\prime \phi_2^\prime
\phi_3^\prime$ of measurement devices.
This condition is well known in the measure theory, namely this is
the condition of equivalence of
two measures: they are absolutely continuous with respect to each
other. The absolute continuity
implies that the transition from one setting of measurement devices
to another is sufficiently smooth
(in measure-theoretical sense). There exists so called Radon-
Nikodim derivative:
\[\frac{{\rm{d}}{\bf P}_{\phi_1 \phi_2 \phi_3}}{{\rm{d}}
{\bf P}_{\phi_1^\prime \phi_2^\prime \phi_3^\prime}}
(\omega)=
{\rm{f}}(\omega;\phi_1 \phi_2\phi_3/\phi_1^\prime
\phi_2^\prime \phi_3^\prime).\] 
The GHZ paradox (via our frequency analysis) demonstrated
that quantum measurement
procedures induce probability distributions which transform
nonsmoothly (in measure-theoretical
sense) from one setting to another. 

Measure-theoretical
singularity is described by the notion of
singularity: ${\bf P}^\prime \perp {\bf P}^{\prime\prime}$ if there is a
set ${\rm{E}}\in {\rm{F}}$ such
that ${\bf P}^{\prime\prime}({\rm{E}})=1$ and ${\bf P}^\prime({\rm{E}})=
0.$ Suppose that ${\bf P}_{\rm{i}}
\perp{\bf P}_{\rm{j}}, {\rm{i,j}}=1,2,\ldots, 4,$ where
${\bf P}_{\rm{j}}$ are probability distributions
in the GHZ scheme. Let $\Omega_{{\rm{j}}}^+, {\rm{j}}=1,2,3,$ play
the role of E in the definition
of ${\bf P}_{\rm{j}}\perp{\bf P}_{4}:{\bf P}_{\rm{j}}(\Omega_{{\rm{j}}}^+)=
1 $ and $ {\bf P}_4(\Omega_{\rm{j}})=
0, {\rm{j}}=1,2,3.$ Then ${\bf P}_4(\Sigma^+)={\bf P}_4(\Omega_{1}
^+\cap\Omega_{2}^+\cap\Omega_{3}^+)=0.$
Thus there is no GHZ paradox. 

We remark that if the space of hidden variables has infinite
dimension, then, for many classes of probability distributions (in
particular, Gaussian), we have equivalence/singularity dichotomy:
either equivalent or singular [15]. It may be that the split of
reality into classical and quantum is just the exhibition of such
a dichotomy.

\section{`Gedanken kollektiven'(counterfactural arguments)}

We note that in the frequency approach the GHZ paradox can be
obtained via counterfactural arguments
(compare with [2], [20]). These arguments are represented here via the
use of 'gedanken kollektiven'. In
fact, the GHZ scheme is applied to four settings $(\pi/2,0,0), (0,
\pi/2,0), (0,0,\pi/2),
(\pi/2,\pi/2,\pi/2).$ Let us consider a `gedanken
kollektiv' corresponding to the simultaneous
imaginary measurement for all angles involved in the GHZ scheme:
$\phi_1=0, \pi/2, \phi_2=0, \pi/2,
\phi_3=0, \pi/2.$ Such an imaginary measurement would be described
by the hidden variable: 
\begin{equation}
\label{q1}
\tilde{\omega}=(\lambda, \lambda_{0}^1, \lambda_{\pi/2}^1, \lambda_{0}
^2, \lambda_{\pi/2}^2, \lambda_{0}^3,
\lambda_{\pi/2}^3).
\end{equation}
Of course, such a measurement is forbidden by
the quantum theory. We recognize
this. However, we continue our frequency analysis trying to find
the origin of the impossibility of
such a measurement. We may image that there are two
settings $\phi_1=0, \pi/2$ for the first photon,
two settings $\phi_2=0, \pi/2$ for the second photon and two
settings $\phi_3 = 0, \pi/2$  for the third photon in the
triple. At the moment of interaction (imaginary) with photons
measurement devices with these
settings have hidden parameters included in (\ref{q1}). If we assume
that the sequence of parameters
$\tilde{\omega}$ corresponding to the sequence of imaginary
measurements, $x=(\tilde{\omega}_{\rm{j}},
{\rm{j}}=1, \ldots, \infty)$ is a collective, then we obtain the
frequency probability distribution
${\bf P}={\bf P}_{\rm{x}}$ which can be used in the GHZ scheme (and
induce the paradox). The origin of the nonexistence of ${\bf P}$
(statistical stabilization in ${\rm{x}}$) is
that collectives corresponding to
incompatible settings of measurement devices are not combinable:
${\rm{x}}_1=(\tilde{{\rm{\omega}}}_{{\rm{j}}}^1), \tilde{{\rm{\omega}}}
_{{\rm{j}}}
^1=(\lambda, \lambda^1, \lambda_{0}^2, \lambda_{0}^3),$ and
${\rm{x}}_2=(\tilde{{\rm{\omega}}}_{{\rm{j}}}^2), \tilde{{\rm{\omega}}}
_{{\rm{j}}}^2=(\lambda, \lambda_{\pi_2}^1,\lambda_{\pi_2}^2,
\lambda_{\pi_2}^3) $ or ${\rm{x}}_1=(\tilde{{\rm{\omega}}}_{{\rm{j}}}
^1), \tilde{{\rm{\omega}}}_{{\rm{j}}}^1=(\lambda, \lambda_{\pi_2}^1,
\lambda_{0}^2, \lambda_{0}^3), $ and ${\rm{x}}_2=(\tilde{{\rm{\omega}}}
_{{\rm{j}}}^2), \tilde{{\rm{\omega}}}_{{\rm{j}}}^2=(\lambda,
\lambda_{0}^1,\lambda_{\pi_2}^2, \lambda_{\pi_2}^3), \ldots $

Thus our frequency counterfactural analysis demonstrated again
that the origin of the GHZ paradox is the existence of
incompatible settings of measurement apparatuses (uncombinable
collectives).

\medskip

{\bf References}

\medskip

[1] J.S. Bell,  Rev. Mod. Phys., {\bf 38}, 447--452 (1966).
J. S. Bell, {\it Speakable and unspeakable in quantum mechanics.}
Cambridge Univ. Press (1987).

[2] J.F. Clauser , M.A. Horne, A. Shimony, R. A. Holt,
Phys. Rev. Letters, {\bf 49}, 1804-1806 (1969);
J.F. Clauser ,  A. Shimony,  Rep. Progr.Phys.,
{\bf 41} 1881-1901 (1978).
A. Aspect,  J. Dalibard,  G. Roger, 
Phys. Rev. Lett., {\bf 49}, 1804-1807 (1982);
D. Home,  F. Selleri, Nuovo Cim. Rivista, {\bf 14},
2--176 (1991). H. P. Stapp, Phys. Rev., D, {\bf 3}, 1303-1320 (1971);
P.H. Eberhard, Il Nuovo Cimento, B, {\bf 38}, N.1, 75-80(1977); Phys. Rev. Letters,
{\bf 49}, 1474-1477 (1982);
A. Peres,  Am. J. of Physics, {\bf 46}, 745-750 (1978).
P. H. Eberhard,  Il Nuovo Cimento, B,
{\bf 46}, N.2, 392-419 (1978); J. Jarrett, Noûs, {\bf 18},
569 (1984).

[3] B. d'Espagnat, {\it Veiled Reality. An anlysis of present-day
quantum mechanical concepts.} Addison-Wesley(1995).
A. Shimony, {\it Search for a naturalistic world view.} Cambridge Univ. Press (1993);
T. Maudlin, {\it Quantum non-locality and relativity.} Blackwill.

[4] L. de Broglie, {\it La thermodynamique de la particule isolee.} Gauthier-Villars, Paris,
1964; G. Lochak, Found. Physics, {\bf 6´}, 173-184 (1976);
E. Nelson, {\it Quantum fluctuation.} Princeton Univ. Press, 1985;
W. De Muynck and W. De Baere W.,
Ann. Israel Phys. Soc., {\bf 12}, 1-22 (1996);
W. De Muynck, W. De Baere, H. Marten,
Found. of Physics, {\bf 24}, 1589--1663 (1994);
W. De Muynck, J.T. Stekelenborg,  Annalen der Physik, {\bf 45},
N.7, 222-234 (1988).
 
[5] E. Beltrametti, G. Cassinelli, {\it The logic of quantum mechanics.} Addison-Wesley, Reading (1981).
 
[6] L. Accardi, {\it Urne e Camaleoni: Dialogo sulla realta,
le leggi del caso e la teoria quantistica.} Il Saggiatore, Rome (1997);
Accardi  L., The probabilistic roots of the quantum mechanical paradoxes.
{\it The wave--particle dualism. A tribute to Louis de Broglie on his 90th 
Birthday}, Edited by S. Diner, D. Fargue, G. Lochak and F. Selleri.
D. Reidel Publ. Company, Dordrecht, 47--55(1984);

[7] I. Pitowsky,  Phys. Rev. Lett, {\bf 48}, N.10, 1299-1302 (1982);
Phys. Rev. D, {\bf 27}, N.10, 2316-2326 (1983);
S.P. Gudder,  J. Math Phys., {\bf 25}, 2397- 2401 (1984);
A. Fine,  Phys. Rev. Letters, {\bf 48}, 291--295 (1982);
P. Rastal, Found. Phys., {\bf 13}, 555 (1983).
W. Muckenheim,  Phys. Reports, {\bf 133}, 338--401 (1986);
W. De Baere,  Lett. Nuovo Cimento, {\bf 39}, 234-238 (1984);
{\bf 25}, 2397- 2401 (1984).

[8] A. Yu. Khrennikov,  Dokl. Akad. Nauk SSSR, ser. Matem.,
{\bf 322}, No. 6, 1075--1079 (1992); J. Math. Phys., {\bf 32}, No. 4, 932--937 (1991);
Physics Letters A, {\bf 200}, 119--223 (1995);
Physica A, {\bf 215}, 577--587 (1995);   Int. J. Theor. Phys., {\bf 34},
2423--2434 (1995);  J. Math. Phys., {\bf 36},
No.12, 6625--6632 (1995);
A.Yu. Khrennikov, {\it $p$-adic valued distributions in 
mathematical physics.} Kluwer Academic Publishers, Dordrecht (1994);
A.Yu. Khrennikov, {\it Non-Archimedean analysis: quantum
paradoxes, dynamical systems and biological models.}
Kluwer Acad.Publ., Dordreht, The Netherlands, 1997;

[9] A. Yu. Khrennikov, {\it Bell and Kolmogorov: probability, reality and nonlocality.}
Reports of Vaxjo Univ., N. 13 (1999); A. Yu. Khrennikov, {\it Interpretations of probability.}
VSP Int. Sc. Publ., Utrecht, 1999.

[10] A. Einstein, B. Podolsky, N. Rosen,  Phys. Rev., {\bf 47}, 777--780
(1935).

[11] D. Greenberger, M. Horne, A. Zeilinger, Going beyond Bell's theorem,
in {\it Bell's theorem, quantum theory, and conceptions of the universe.} Ed. M.Kafatos,
Kluwer Academic, Dordrecht, 73-76 (1989).

[12] A. N. Kolmogoroff, {\it Grundbegriffe der Wahrscheinlichkeitsrech}
Springer Verlag, Berlin (1933); reprinted:
{\it Foundations of the Probability Theory}. 
Chelsea Publ. Comp., New York (1956).

[13] R. von Mises, Grundlagen der Wahrscheinlichkeitsrechnung.\linebreak 
{\it Math.Z.,} {\bf 5}, 52--99 (1919);
R. von Mises, {\it Probability, Statistics and Truth},
 Macmillan, London (1957).

[14] R.  von Mises, {\it The mathematical theory of probability and
 statistics}. Academic, London (1964);  E. Tornier , {\it Wahrscheinlichkeitsrechnunug und allgemeine 
Integrationstheorie.}, Univ. Press, Leipzing (1936);

[15] A. N. Shiryayev, {\it Probability.} Springer, New York-Berlin-Hei (1984).

[16]  E. Kamke, \"Uber neuere Begr\"undungen der 
Wahrscheinlichkeitsrechnung. {\it Jahresbericht der Deutschen Matemati,}
{\bf 42,} 14-27 (1932);
J. Ville, {\it Etude critique de la notion de collective}, Gauthier--
Villars, Paris (1939);M. Van Lambalgen, Von Mises' definition of random sequences
reconsidered. {\it J. of Symbolic Logic,} {\bf 52,} N. 3 (1987).

[17] A. Wald, Die Widerspruchsfreiheit des Kollektivbegriffs
in der Wahrscheinlichkeitsrechnung. {\it Ergebnisse eines Math. Kolloquiums},
{\bf 8}, 38-72 (1938).

[18] Venn J., {\it The logic of chance.} London (1866);
reprint: Chelsea, New York (1962).

[19] L. E. Ballentine,  Rev. Mod. Phys., {\bf 42}, 358--381 (1970).

[20] I. Kvart, {\it A theory of Counterfactuals.} Indiapolis: Hackett(1986);
D. Lewis, {\it Counterfactuals.} Cambridge, Mass.: Harvard Univ. Press (1973);

\end{document}